\documentclass{article}
\usepackage{spconf}
\usepackage{amsmath,amssymb,amsthm}
\usepackage{enumerate}
\usepackage{stfloats}
\usepackage{comment}
\usepackage{subfig}
\usepackage{amsfonts}
\usepackage[utf8]{inputenc}
\usepackage[english]{babel}
\usepackage[dvipsnames]{xcolor}
\definecolor{myPink}{RGB}{255,105,183}
\usepackage[margin=0.9in]{geometry}
\usepackage[numbers,sort&compress]{natbib}

\usepackage[T1]{fontenc}
\usepackage{graphics} 
\usepackage{epsfig} 
\usepackage[mathscr]{euscript}
\usepackage{algorithm}
\usepackage[noend]{algpseudocode}
\makeatletter
\def\BState{\State\hskip-\ALG@thistlm}
\makeatother

\usepackage{hyperref}

\usepackage{tikz}
\usetikzlibrary{arrows,shapes,chains,matrix,positioning,scopes,patterns,calc,
decorations.markings,
decorations.pathmorphing,
}

\usepackage{pgfplots}
\pgfplotsset{compat=1.3}
\usepgflibrary{shapes}

\renewcommand{\epsilon}{\varepsilon}

\newcommand{\RNum}[1]{\uppercase\expandafter{\romannumeral #1\relax}}

\newcommand{\hv}{\mathrm{\mathbf{h}}}
\newcommand{\yv}{\mathrm{\mathbf{y}}}
\newcommand{\wv}{\mathrm{\mathbf{w}}}
\newcommand{\xv}{\mathrm{\mathbf{x}}}
\newcommand{\sv}{\mathrm{\mathbf{s}}}
\newcommand{\zv}{\mathrm{\mathbf{z}}}

\def\Pr{\mathrm{Pr}}

\DeclareMathAlphabet{\mcl}{OMS}{cmsy}{m}{n}

\newlength\tikzwidth
\newlength\tikzheight

\definecolor{mycolor1}{rgb}{0.63529,0.07843,0.18431}%
\definecolor{mycolor2}{rgb}{0.00000,0.44706,0.74118}%
\definecolor{mycolor3}{rgb}{0.00000,0.49804,0.00000}%
\definecolor{mycolor4}{rgb}{0.87059,0.49020,0.00000}%
\definecolor{mycolor5}{rgb}{0.00000,0.44700,0.74100}%
\definecolor{mycolor6}{rgb}{0.74902,0.00000,0.74902}%

\def\fig_path{./Figures}
\newif\ifproof
\prooffalse

\title{Asynchronous Neighbor Discovery\\Using Coupled Compressive Sensing}
\name{\begin{tabular}{c} \dag Vamsi K. Amalladinne,
\dag Krishna R. Narayanan, \\
\dag Jean-Francois Chamberland,
	\S Dongning Guo \end{tabular}
\thanks{
This material is based upon work supported by the National Science Foundation (NSF) under Grant No.~CCF-1619085.}
}
\address{\dag Department of Electrical and Computer Engineering, Texas A\&M University \\
\S Department of Electrical Engineering and Computer Science, Northwestern University}
\begin{document}

\maketitle

\begin{abstract}
The neighbor discovery paradigm finds wide application in Internet of Things networks, where the number of active devices is orders of magnitude smaller than the total device population.
Designing low-complexity schemes for asynchronous neighbor discovery has recently gained significant attention from the research community.
Concurrently, a divide-and-conquer framework, referred to as coupled compressive sensing, has been introduced for the synchronous massive random access channel. 
This work adapts this novel algorithm to the problem of asynchronous neighbor discovery with unknown transmission delays.
Simulation results suggest that the proposed scheme requires much fewer transmissions to achieve a performance level akin to that of state-of-the-art techniques.
\end{abstract}


\section{Introduction}
\label{sec:intro}

Neighbor discovery is a frequently occurring task in Internet of Things (IoT) networks \cite{chen2017sparse}. In this context, a node within a wireless network seeks to identify all neighbors with which it can communicate directly. The total number of potential devices in a typical IoT network is on the order of billions. Hence, there is a pressing need to design algorithms whose computational complexity scales moderately with the number of devices and bandwidth \cite{vasudevan2009neighbor,borbash2007asynchronous,angelosante2010neighbor,zhang2013neighbor, thompson2018compressed, applebaum2012asynchronous, chen2017sparse}. Further, because devices are at different geographical locations, their transmissions are typically asynchronous as perceived by the receiver. It is thus important to design coding schemes that perform well in the presence of asynchronous transmissions.

The interest in finding good solution to neighbor discovery is evinced by the large number of contributions on this topic.
In particular, tools from compressive sensing have been employed to design algorithms for neighbor discovery in \cite{zhang2013neighbor, thompson2018compressed, applebaum2012asynchronous, chen2017sparse}. Two different paradigms are proposed for synchronous neighbor discovery in~\cite{zhang2013neighbor}. The first algorithm employs random on-off signatures and the query node discovers its neighbors using a variation of group testing. The other algorithm is based on second order Reed-Muller codes followed by chirp reconstruction. In~\cite{thompson2018compressed}, the authors employ sparse Kerdock matrices as codebooks for synchronous neighbor discovery. Still, the schemes developed in \cite{zhang2013neighbor, thompson2018compressed} rely on symbol synchronicity across transmissions, and they cannot be used when signals are asynchronous. In~\cite{applebaum2012asynchronous}, LASSO is employed for node identification in an asynchronous random access scenario. Regrettably, the schemes mentioned above do not scale well when the device address space becomes huge (e.g., 48-bit IEEE 802.11 MAC addresses \cite{ieee1999wireless}). In~\cite{chen2017sparse}, an asynchronous neighbor discovery scheme based on sparse orthogonal frequency division multiplexing (S-OFDM) is proposed. This approach is shown to scale well with the device address space. That is, the computational complexity of this scheme scales linearly with the number of active devices, and logarithmically with the cardinality of the address space. 

In this paper, we develop a scalable, asynchronous neighbor discovery algorithm tailored to large address spaces. This novel approach requires substantially fewer transmissions (smaller bandwidth) than previously published alternatives~\cite{chen2017sparse} to achieve a comparable error performance. The proposed scheme builds on our earlier work on coupled compressive sensing (CCS), which first appeared in the treatment of the unsourced and uncoordinated multiple access channel~\cite{amalladinne2018coupled}.
Herein, we demonstrate that a variant of the CCS algorithm outperforms state-of-the-art schemes for asynchronous neighbor discovery as well.  There are a few key features that distinguish the massive random access paradigm considered in \cite{amalladinne2018coupled} and the neighbor discovery problem at hand. Only additive white Gaussian noise (AWGN) appears in~\cite{amalladinne2018coupled}, whereas the present article considers a flat fading environment where channel coefficients are unknown at the receiver. Also, transmissions are synchronous in~\cite{amalladinne2018coupled}; the modified CCS algorithm must thence be extended to asynchronous cases. 

Having reviewed existing contributions, we turn to the presentation of our scheme.
Throughout, we employ the following notation. $\mathbf{DFT}_k$ symbolizes the $k \times k$ discrete-Fourier transform matrix. The cardinality of set $\mathcal{S}$ is represented by $|\mathcal{S}|$. We employ $\mathbb{C}$, $\mathbb{Z}$, and $\mathbb{Z}_+$ to denote complex numbers, integers, and non-negative integers, respectively. For any $a,b \in \mathbb{Z}$, $[a:b] = \{ c \in \mathbb{Z}:a \le c \le b\}$. For a vector $\xv= [ x_1 ~ x_2 \cdots x_n ] \in \mathbb{C}^n$, $\operatorname{supp}(\xv) = \{k \in [1:n] : x_k \neq 0\}$, $\| \xv \|_0 = | \operatorname{supp}(\xv) |$, and $\| \xv \|_p$ is the standard $\ell_p$ norm with $p \ge 1$. We write $X \sim \mathbb{U}[a,b]$ to indicate that $X$ is uniformly distributed between $a$ and $b$. Symbols $\mathbf{0}_{n}$, $\mathbf{0}_{k\times n}$ denote the all-zero column vector of length $n$ and the $k \times n$ all-zero matrix, respectively. We use $\delta_{i,j}$ to denote the standard Kronecker delta function.

\section{System Model}
\label{sec:systemmodel}

Let $\mathcal{K}_{\mathrm{tot}}$ be the collection of all nodes and $\mathcal{K} \subseteq \mathcal{K}_{\mathrm{tot}}$ be the subset of active nodes within a network.
For convenience, we assign parameters to the cardinalities of these sets, namely $|\mathcal{K}_{\mathrm{tot}}| = K_{\mathrm{tot}}$ and $|\mathcal{K}| = K$. An active device $k \in \mathcal{K}$ transmits a frame (codeword) $\xv_k=[x_{k,0}~ x_{k,1} \cdots x_{k,N-1}]^\mathrm{T} \in \mathbb{C}^N$ of $N$ complex symbols with $\mathbb{E}[ \| \xv_{k} \|_2^2]=N$. The signal received by the query node at time instant~$i$ is given by
\begin{equation}
y_i = \sum_{k \in \mathcal{K}} \sqrt{P} h_k x_{k,i-\tau_k} + w_i, \label{yi}
\end{equation}
where $h_k$ denotes the complex fading coefficient of the link between node~$k$ and the query node, $\tau_k$ is the discrete delay in terms of symbols experienced by the signal sent by device~$k$, $P$ represents the transmission power, and $w_i$ is complex additive white Gaussian noise with zero mean and unit variance. We explore two fading models.

\noindent
\textbf{\textit{Fading Model I:}} To facilitate a fair comparison between the proposed scheme and the approach of~\cite{chen2017sparse}, we assume that the magnitude of each fading coefficient is bounded above and below by two constants, $\underline{h}$ and $2\underline{h}$, and its phase is distributed uniform between $0$ and $2\pi$. Specifically, $|h_k| \sim \mathbb{U}[\underline{h}, 2\underline{h}]$ and $\angle{h_k} \sim \mathbb{U}[0,2\pi]$ for all $k \in \mathcal{K}$. The signal-to-noise ratio (SNR) for this model is equal to $\mathsf{SNR} = {P|\underline{h}|^2}/{2}$.

\noindent
\textbf{\textit{Fading Model II:}} In this model, we incorporate the effects of network topology and signal propagation loss into fading. Similar to \cite{zhang2013neighbor}, we assume that the channel gains follow a Pareto distribution with scale parameter $\eta > 0$ and shape parameter $\alpha > 1$\footnotemark. The transmission SNR for this model is given by $\mathsf{SNR} = {P\eta}/{2}$.
\footnotetext{See \cite{zhang2013neighbor} for details regarding how this model is relevant to the neighbor discovery problem.}

The delays are specific to each communication link, and remain constant over the duration of a frame. Further, they are bounded above by a known constant $\mathcal{T}$, i.e., $\mathcal{T} = \max_{k \in \mathcal{K}} \tau_k$, since we are interested only in discovering neighbors who fall within a bounded distance. We refer to the special case $\mathcal{T}=0$ as synchronous neighbor discovery. The receiver assumes that signals are only present during the frame duration, i.e., $s_{k,i} = 0$ when $i<0$ or $i>N-1$  for $k \in \mathcal{K}$.
The receiver is tasked with finding $\hat{\mathcal{K}}$, an estimate of the unknown collection $\mathcal{K}$ of active nodes, given measurements $\yv=[y_0~y_1~\cdots ~y_{N-1}]^{\mathrm{T}}$. To facilitate neighbor discovery, every active device needs to embed its identity in the transmission. Since the system features $K_{\mathrm{tot}}$ devices, nodes can be distinguished using roughly $\log_2 K_{\mathrm{tot}}$ bits, which are encoded into $N$ complex symbols and sent. Thus, we have a multiple access system, with $K$ active devices intending to transmit $\log_2 K_{\mathrm{tot}}$ bits each by encoding them into $N$ complex symbols.

The error probability of the system is defined as the probability that the estimated set $\hat{\mathcal{K}}$ is not the same as the actual set $\mathcal{K}$ of active devices.
Mathematically, $P_{\mathrm{e}} = \Pr [\hat{\mathcal{K}} \neq \mathcal{K}]$. Our objective is to design a low-complexity coding scheme that offers good error performance with respect to SNR at low codeword lengths.


\section{Proposed Scheme}

Our new scheme is based on adapting the CCS algorithm~\cite{amalladinne2018coupled} to the needs of neighbor discovery.
We initiate the discussion with the encoding scheme.
As mentioned above, the amount of information needed to identify a device is $B=\log_2 K_{\mathrm{tot}}$, and we refer to this data as the message bits. Every $B$-bit message is encoded into $N$ complex symbols, and active devices concurrently transmit the resulting signals. The encoding process features an outer encoder, which is referred to as the tree encoder, and an inner CS encoder; these two parts are explained below.

\noindent
\textbf{\textit{Tree Encoder:}}
The tree encoder splits the $B$-bit identity into several sub-blocks, each of which is amenable to CS recovery. To this end, the $B$-bit message is partitioned into $n$ sub-blocks and redundancy is added to some blocks. Let $m_i$ denote the number of message bits present in the $i^{\mathrm{th}}$ sub-block, $\sum_{i=0}^{n-1}m_i=B$. Let $l_i$ denote the number of redundant parity-check bits added to sub-block~$i$, with $l_0=0$; We never add parity bits to the $0^{\mathrm{th}}$ sub-block. The length of each coded sub-block is $J$, i.e., $m_i + l_i = J$ for $i \in [0:n-1]$. We denote the length of coded block by $M$, which yields $B + \sum_{i=0}^{n-1}l_i = M$. The parity-check bits added in the $i^{\mathrm{th}}$ sub-block act as random parity check constraints for all the message bits preceding it\footnotemark.
\footnotetext{See \cite{amalladinne2018coupled} for more details regarding the construction of parity-check bits, their allocation across sub-blocks and the theoretical analysis of error probability and computational complexity of the CCS scheme.}

\noindent
\textbf{\textit{CS Encoder:}}
Every coded sub-block of $J$ bits is further encoded into $\tilde{N}=\frac{N}{n}$ complex symbols by the CS encoder. That is, every $M$-bit output of the tree encoder is mapped into $N$ complex symbols by the CS encoder. Specifically, each sub-block corresponding to a device is first encoded into $\tilde{N}-\mathcal{T}$ complex symbols via a matrix $\mathbf{S}=[\sv_1 ~\sv_2 \cdots \sv_{2^J}]\in \mathbb{C}^{(\tilde{N}-\mathcal{T})\times 2^J}$. This is accomplished through a bijective function $f:\{0,1\}^J \mapsto \{\sv_j, j \in [1:2^J]\}$ that maps sub-blocks into columns of $\mathbf{S}$. When transmissions are synchronous ($\mathcal{T}=0$), the receiver observes superimposed sub-blocks and it can recover individual sub-blocks by implementing a CS decoding algorithm. On the other hand, when transmissions are asynchronous, there may be interference from other sub-blocks. To prevent this from occurring, each column in $\mathbf{S}$ is padded with $\mathcal{T}$ zeros to form the codebook $\mathbf{S}_{\mathrm{ZP}}=[ {\mathbf{S}}^{\mathrm{T}} \mathbf{0}_{\mathcal{T} \times 2^J}^{\mathrm{T}}]^{\mathrm{T}} \in \mathbb{C}^{\tilde{N}\times 2^J}$. In other words, each sub-block of length $J$ bits is encoded into a column of $\mathbf{S}_{\mathrm{ZP}}$ and transmitted over the channel.

At this stage, we turn to the decoding process, which also features two components: an inner CS decoder, which recovers sub-blocks from asynchronous noisy measurements; and an outer tree decoder, which connects all the sub-blocks corresponding to a parent message.

\noindent
\textbf{\textit{CS decoder:}}
A compressive sensing framework for asynchronous multiple access was first developed in~\cite{applebaum2012asynchronous}. Let $\mathcal{I}^{(i)}=\{i_k:~k \in K\} \subset [1:2^J]$ denote the indices corresponding to the columns of $\mathbf{S}_{\mathrm{ZP}}$ transmitted during $i^{\mathrm{th}}$ sub-block. Since the maximum delay $\mathcal{T}$ is known to the decoder, the signal received during the $i^{\mathrm{th}}$ sub-block can be written as
\begin{equation}
\yv^{(i)} = {\textstyle \sum_{k \in \mathcal{K}}}
h_k\tilde{\sv}^{(\tau_k)}_{i_k}+\wv^{(i)}, \label{asynchronousCS1}
\end{equation}
where $\tilde{\sv}^{(\tau_k)}_{i_k}=[\mathbf{0}^\mathrm{T}_{\tau_k}~\sv_{i_k}^{\mathrm{T}}~\mathbf{0}^\mathrm{T}_{\mathcal{T}-\tau_k}]^\mathrm{T} \in \mathbb{C}^{\tilde{N}}$,
a shifted version of $\sv_{i_k}^{\mathrm{T}}$ with zero paddings. Equation~\eqref{asynchronousCS1} can be expressed in matrix form as
\begin{equation}
\yv^{(i)} = \tilde{\mathbf{S}}\tilde{\hv}^{(i)} + \wv^{(i)}, \label{asycnronousCS}
\end{equation}
where $\tilde{\mathbf{S}}=[\tilde{\mathbf{S}}_1~\tilde{\mathbf{S}}_2 \cdots \tilde{\mathbf{S}}_{2^J}]$ is an ${\tilde{N} \times 2^J(\mathcal{T}+1)}$ matrix, and
sub-matrices $\tilde{\mathbf{S}}_j=[\tilde{\sv}^{(0)}_j~\tilde{\sv}^{(1)}_j \cdots \tilde{\sv}^{(\mathcal{T})}_j]
\in \mathbb{C}^{\tilde{N}\times (\mathcal{T}+1)}$ for $j \in [1:2^J]$.
The vectors $\tilde{\hv}^{(i)} = [\tilde{\hv}_1^{(i)}~\tilde{\hv}_2^{(i)} \cdots \tilde{\hv}_{2^J}^{(i)}]^T
\in \mathbb{C}^{2^J(\mathcal{T}+1)}$ are defined by
\begin{equation}
\tilde{\hv}_j^{(i)}= {\textstyle \sum_{k \in \mathcal{K}}} [\mathbf{0}^\mathrm{T}_{\tau_k}~h_k~\mathbf{0}^\mathrm{T}_{\mathcal{T}-\tau_k}]^\mathrm{T} \delta_{j,{i_k}}.
\end{equation}
Equation~\eqref{asycnronousCS} resembles a standard compressed sensing setting with $\yv^{(i)}$ acting as the observation vector and $\tilde{\mathbf{S}}$ being the sensing matrix. The CS decoder constructs the sensing matrix $\tilde{\mathbf{S}}$ using the procedure outlined above. An estimate $\hat{\tilde{\hv}}^{(i)}$ of vector $\tilde{\hv}^{(i)}$ is first computed using measurements $\yv^{(i)}$ by implementing any standard CS decoding technique. The final estimate of $\tilde{\hv}^{(i)}$ is subsequently obtained by taking the best $K$-term approximation{\footnotemark}\ to the vector $\hat{\tilde{\hv}}^{(i)}$.
\footnotetext{The best $s$-term approximation to a vector $\xv \in \mathbb{C}^{n}$ is defined as
$\hat{\xv}_s = {\textstyle \operatorname*{argmin}_{\zv \in \mathbb{C}^n, ||\zv||_0 \le s}} ||\xv-\zv||_2.$}

\noindent
\textbf{\textit{Tree Decoder:}}
The purpose of the tree decoder is to string together the message components identified by the CS decoder.
Our choice algorithm is adapted from~\cite{amalladinne2018coupled}.
Yet, in contrast to our previous work, the concatenation process is informed by both the parity-check bits and the fading coefficients returned by the CS decoder.
Integrating the soft estimates is a means to reduce the amount of redundancy needed in the tree code.
Pragmatically, the ability to discriminate between fade levels is key in successfully leveraging soft estimates.
When the fading profile is severe, the coefficients become very useful in terms of piecing together message components.
However, when fade variations are minimal, the system must revert back to adding parity bits.
The tree decoding algorithm is explained below.

\begin{figure}[!ht]
\centering
\resizebox{0.4\textwidth}{!}{\usetikzlibrary{arrows}
 \usetikzlibrary{matrix}
\begin{tikzpicture}

\draw  [ultra thick,fill= gray](0.5,0.5) ellipse (0.3 and 0.3);
\draw  [ultra thick,fill=gray!45](-2,-1.5) ellipse (0.3 and 0.3);
\draw  [ultra thick,fill=gray](-0.5,-1.5) ellipse (0.3 and 0.3);
\draw  [ultra thick](3,-1.5) ellipse (0.3 and 0.3);
\node at (0.65,-1.5) {\huge\textbf{.}};
\node at (1.3,-1.5) {\huge\textbf{.}};
\node at (1.9,-1.5) {\huge\textbf{.}};
\node (v1) at (0.3,0.4) {};
\node (v2) at (-1.8,-1.35) {};
\draw [ultra thick][-triangle 90] (v1) edge (v2);
\node (v3) at (0.4,0.3) {};
\node (v4) at (-0.35,-1.3) {};
\draw [ultra thick][-triangle 90] (v3) edge (v4);
\node (v6) at (0.65,0.4) {};
\node (v5) at (2.8,-1.35) {};
\draw [ultra thick][-triangle 90] (v6) edge (v5);

\draw [ultra thick] (-4.5,-3.5) ellipse (0.3 and 0.3);
\draw [ultra thick] (-3.8,-3.5) ellipse (0.3 and 0.3);
\draw [ultra thick] (-2.5,-3.5) ellipse (0.3 and 0.3);
\draw [ultra thick,fill=gray!45] (-1,-3.5) ellipse (0.3 and 0.3);
\draw [ultra thick] (-0.3,-3.5) ellipse (0.3 and 0.3);
\draw [ultra thick,fill=gray] (1,-3.5) ellipse (0.3 and 0.3);
\draw [ultra thick] (3,-3.5) ellipse (0.3 and 0.3);
\draw [ultra thick] (3.7,-3.5) ellipse (0.3 and 0.3);
\draw [ultra thick] (5,-3.5) ellipse (0.3 and 0.3);
\node at (-3.3,-3.5) {\huge\textbf{.}};
\node at (-2.9,-3.5) {\huge\textbf{.}};
\node at (0.2,-3.5) {\huge\textbf{.}};
\node at (0.55,-3.5) {\huge\textbf{.}};
\node at (1.75,-3.5) {\huge\textbf{.}};
\node at (2,-3.5) {\huge\textbf{.}};
\node at (2.3,-3.5) {\huge\textbf{.}};
\node at (4.15,-3.5) {\huge\textbf{.}};
\node at (4.5,-3.5) {\huge\textbf{.}};
\node (v8) at (-2.4,-1.75) {};
\node (v7) at (-4.35,-3.35) {};
\draw [ultra thick][-triangle 90] (v8) edge (v7);
\node (v9) at (-2.1,-1.9) {};
\node (v10) at (-2.4,-3.25) {};
\draw [ultra thick][-triangle 90] (v9) edge (v10);
\node (v11) at (-0.65,-1.8) {};
\node (v12) at (-0.9,-3.25) {};
\draw [ultra thick][-triangle 90] (v11) edge (v12);
\node (v13) at (-0.25,-1.8) {};
\node (v14) at (0.9,-3.3) {};
\draw [ultra thick][-triangle 90] (v13) edge (v14);
\node (v15) at (2.95,-1.8) {};
\node (v16) at (3,-3.2) {};
\draw [ultra thick][-triangle 90] (v15) edge (v16);
\node (v17) at (3.35,-1.8) {};
\node (v18) at (4.9,-3.3) {};
\draw [ultra thick][-triangle 90] (v17) edge (v18);
\node (v20) at (-3.65,-3.35) {};
\node (v19) at (-2.2,-1.85) {};
\draw [ultra thick][-triangle 90] (v19) edge (v20);
\node (v21) at (-0.45,-1.8) {};
\node at (-0.5,-3) {};
\node (v22) at (-0.4,-3.25) {};
\draw [ultra thick][-triangle 90] (v21) edge (v22);
\node (v23) at (3.1,-1.8) {};
\node (v24) at (3.6,-3.3) {};
\draw [ultra thick][-triangle 90] (v23) edge (v24);
\node (v25) at (0.5,0.5) {};
\node at (-2,-1.5) {};
\node (v26) at (-0.5,-1.5) {};
\node at (3,-1.5) {};
\node at (-4.5,-3.5) {};
\node at (-3.75,-3.5) {};
\node at (-2.5,-3.5) {};
\node at (-1,-3.5) {};
\node at (-0.3,-3.5) {};
\node at (1,-3.5) {};
\node at (3,-3.5) {};
\node at (3.7,-3.5) {};
\node at (5,-3.5) {};
\node at (-4.5,-4) {\huge\textbf{.}};
\node at (-4.5,-4.5) {\huge\textbf{.}};
\node at (-4.5,-5) {\huge\textbf{.}};
\node at (-4.5,-5) {\huge\textbf{.}};
\node at (-4.5,-5) {\huge\textbf{.}};
\node at (-4.5,-5.5) {\huge\textbf{.}};
\node at (-3.8,-4.5) {\huge\textbf{.}};
\node at (-3.8,-4) {\huge\textbf{.}};
\node at (-3.8,-5) {\huge\textbf{.}};
\node at (-3.8,-5.5) {\huge\textbf{.}};
\node at (-2.5,-4) {\huge\textbf{.}};
\node at (-2.5,-4.5) {\huge\textbf{.}};
\node at (-2.5,-5) {\huge\textbf{.}};
\node at (-2.5,-5.5) {\huge\textbf{.}};
\node at (-1,-4) {\huge\textbf{.}};
\node at (-1,-4.5) {\huge\textbf{.}};
\node at (-1,-5) {\huge\textbf{.}};
\node at (-1,-5.5) {\huge\textbf{.}};
\node at (-0.3,-4) {\huge\textbf{.}};
\node at (-0.3,-4.5) {\huge\textbf{.}};
\node at (-0.3,-5) {\huge\textbf{.}};
\node at (-0.3,-5.5) {\huge\textbf{.}};
\node at (1,-4) {\huge\textbf{.}};
\node at (1,-4.5) {\huge\textbf{.}};
\node at (1,-5) {\huge\textbf{.}};
\node at (1,-5.5) {\huge\textbf{.}};
\node at (3,-4) {\huge\textbf{.}};
\node at (3,-4.5) {\huge\textbf{.}};
\node at (3,-5) {\huge\textbf{.}};
\node at (3,-5.5) {\huge\textbf{.}};
\node at (3.7,-4) {\huge\textbf{.}};
\node at (3.7,-4.5) {\huge\textbf{.}};
\node at (3.7,-5) {\huge\textbf{.}};
\node at (3.7,-5.5) {\huge\textbf{.}};
\node at (5,-4) {\huge\textbf{.}};
\node at (5,-4.5) {\huge\textbf{.}};
\node at (5,-5) {\huge\textbf{.}};
\node at (5,-5.5) {\huge\textbf{.}};
\node at (5,-5.5) {\huge\textbf{.}};
\draw [ultra thick] (-5,-6.5) ellipse (0.3 and 0.3);
\draw [ultra thick] (-4.3,-6.5) ellipse (0.3 and 0.3);
\draw [ultra thick] (-3.6,-6.5) ellipse (0.3 and 0.3);
\draw [ultra thick] (-2.9,-6.5) ellipse (0.3 and 0.3);
\draw [ultra thick] (-2.2,-6.5) ellipse (0.3 and 0.3);
\draw [ultra thick] (-1.5,-6.5) ellipse (0.3 and 0.3);
\draw [ultra thick,fill=gray] (-0.8,-6.5) ellipse (0.3 and 0.3);
\draw [ultra thick] (1.8,-6.5) ellipse (0.3 and 0.3);
\draw [ultra thick] (2.5,-6.5) ellipse (0.3 and 0.3);
\draw [ultra thick] (3.2,-6.5) ellipse (0.3 and 0.3);
\draw [ultra thick] (3.9,-6.5) ellipse (0.3 and 0.3);
\draw [ultra thick] (4.6,-6.5) ellipse (0.3 and 0.3);
\draw [ultra thick] (5.3,-6.5) ellipse (0.3 and 0.3);
\draw [ultra thick] (6,-6.5) ellipse (0.3 and 0.3);
\node at (0,-6.5) {\huge\textbf{.}};
\node at (0.5,-6.5) {\huge\textbf{.}};
\node at (1,-6.5) {\huge\textbf{.}};
\node at (0.5,0.5) {};
\node at (-0.25,-1.25) {};
\node (v27) at (1,-3.5) {};
\node (v28) at (-0.8,-6.5) {};
\draw [dotted,line width=0.8mm,color=red] plot[smooth, tension=.8] coordinates {(v3) (v26) (v27) (v28)};
\end{tikzpicture}}
\caption{This graph illustrates the tree decoding process. Fading coefficients are captured by the shade of nodes. At every stage, only nodes that are close to the fade estimate are retained. Parity check constraints are then verified for residual partial paths to further prune the tree. The highlighted path represent the unique survivor of this process, and thereby forms the basis for the decoded message.}
\label{fig:tree diagram}
\end{figure}
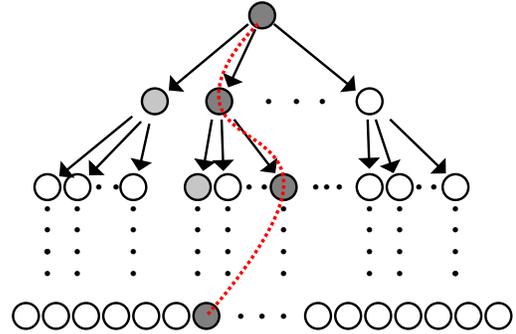

Every candidate sub-block listed by the CS decoder at the onset of the process becomes a root of a decoding tree.
Given a root sub-block, layer~1 of the tree is composed of all
$K$ sub-blocks found on the second list of the CS decoder. This process continues, adding $K$ branches to every node, until the final CS decoding stage is reached.
This results in a tree with $K^{n-1}$ leafs.
The tree decoder then proceeds iteratively to aggregate the sub-blocks that correspond to one parent message among these $K^{n-1}$ possible paths. The vectors $\{\tilde{\hv}^{(i)}\}_{i=0}^{n-1}$ in \eqref{asycnronousCS} have different support sets, yet they share the same non-zero entries because the fading coefficient corresponding to a device does not change across sub-blocks. Thus, the soft fade estimates offered by the CS decoder can act as statistical features to piece together the sub-blocks associated with one parent message. To this end, at stage~$i$ of the tree decoding process, we only retain a subset $\mathcal{L}_i \subseteq [1:K^i]$ of nodes whose estimates of the fading coefficients are comparable to the estimate of the fade level based on the paths retained up to stage~$i-1$. We further prune the decoding tree by only keeping partial paths that satisfy the $l_i$ constraints associated with parity check bits. At stage $i+1$, the tree decoder focuses exclusively on the children of nodes that survived stage~$i$. 
If there exists a single valid path at the end, the decoder outputs the corresponding message; otherwise, it declares a failure. 
Fig.~\ref{fig:tree diagram} illustrates various stages involved in our tree decoding algorithm.

\section{Simulation Results}
\label{sec:sims}

In this section, we present simulation results that showcase the performance of the proposed scheme and compare it to the performance of the scheme found in~\cite{chen2017sparse}. A random sub-matrix of $\mathbf{DFT}_{2^J}$ of dimension $\tilde{N}-\mathcal{T} \times 2^J$ is chosen as the device codebook $\mathbf{S}$ for all simulations. These matrices are known to satisfy the restricted isometry property (RIP) and to be a good choice for noisy compressed sensing~\cite{haviv2017restricted}.
The LASSO algorithm is utilized to solve the CS sub-problems.

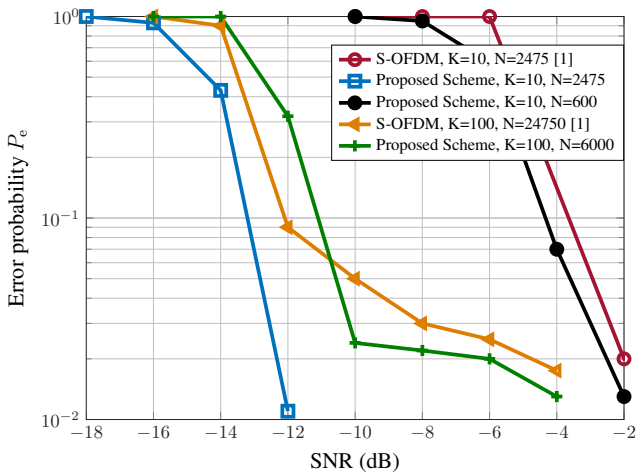
\begin{figure}[bth]
\centering
\resizebox{0.5\textwidth}{!}{\begin{tikzpicture}
\def\fsize{\large}
\pgfplotsset{every y tick label/.append style={font=\fsize}}
\pgfplotsset{every x tick label/.append style={font=\fsize}}

\begin{semilogyaxis}[%
width=4in,
height=3in,
scale only axis,
every outer x axis line/.append style={white!15!black},
every x tick label/.append style={font=\color{white!15!black}},
xmin=-18,
xmax=-2,
xtick = {-18,-16,...,-2},
xlabel={\fsize SNR (dB)},
xmajorgrids,
every outer y axis line/.append style={white!15!black},
every y tick label/.append style={font=\color{white!15!black}},
ymin=0.01,
ymax=1,
yminorgrids = true,
ylabel={\fsize Error probability $P_{\mathrm{e}}$},
ymajorgrids,
legend style={at={(1,0.93)},anchor=north east, draw=black,fill=white,legend cell align=left,font=\small}
]

\addplot [color=mycolor1,solid,line width=2.0pt,mark size=3pt,mark=o,mark options={solid}]
  table[row sep=crcr]{
-10	1\\
-8	1\\
-6	1\\
-2	0.02\\
0	0.023\\
};
\addlegendentry{S-OFDM, K=10, N=2475 \cite{chen2017sparse}};

\addplot [color=mycolor2,solid,line width=2.0pt,mark size=3pt,mark=square,mark options={solid}]
  table[row sep=crcr]{
-20	1\\
-18	1\\
-16	0.93\\
-14	0.43\\
-12	0.011\\
};
\addlegendentry{Proposed Scheme, K=10, N=2475};

\addplot [color=black,solid,line width=2.0pt,mark size=3pt,mark=*,mark options={solid}]
  table[row sep=crcr]{
-10	1\\
-8	0.95\\
-6	0.59\\
-4	0.07\\
-2	0.013\\
0	0.006\\
};
\addlegendentry{Proposed Scheme, K=10, N=600};

\addplot [color=mycolor4,solid,line width=2.0pt,mark size=3pt,mark=triangle,mark options={solid,rotate=90}]
  table[row sep=crcr]{
-16 1\\
-14 0.9\\
-12 0.09\\
-10	0.05\\
-8	0.03\\
-6	0.025\\
-4	0.0175\\
};
\addlegendentry{S-OFDM, K=100, N=24750 \cite{chen2017sparse}};

\addplot [color=mycolor3,solid,line width=2.0pt,mark size=3pt,mark=+,mark options={solid}]
  table[row sep=crcr]{
-16 1\\
-14 1\\
-12 0.32\\
-10	0.024\\
-8	0.022\\
-6	0.02\\
-4	0.013\\
};
\addlegendentry{Proposed Scheme, K=100, N=6000};

\end{semilogyaxis}
\end{tikzpicture}%
}
\caption{This graph shows the error probability of synchronous neighbor discovery as a function of SNR, and hints at the superior performance of the proposed scheme.}
\label{fig:sim_results1}
\end{figure}

\noindent
\textbf{\textit{Fading Model I:}}
We consider a network with $K_{\mathrm{tot}}=2^{38}$ devices and an active device population with $K \in \{10,100\}$. The number of sub-blocks, the length of each sub-block, and the number of parity check bits in each sub-block are chosen as $n=6$, $J=10$, $[l_0~l_1 \cdots l_5]=[0~ 0~ 0~ 2~ 10~ 10]$ for $K=10$; and $n=10$, $J=11$, $[l_0~l_1 \cdots l_9]=[0~ 5~ 7~ 7~ 7~ 7~ 7~ 10~ 11~ 11]$ for $K=100$. For this first fading model, the channel gains are distributed within a fairly small range. As such, the reconstruction error of the soft channel estimates will largely be uninformative. Thus, in this case, tree decoding only relies on parity-check bits to stitch the sub-blocks of a message together. In Fig.~\ref{fig:sim_results1}, the error probability of synchronous neighbor discovery is plotted as a function of SNR. For the same number of channel uses, the proposed scheme provides about 10~dB improvement over the results in~\cite{chen2017sparse}.
Similarly, for a same error performance, the required codeword length is reduced by approximately 75\%.

\begin{figure}[tbh]
\centering
\resizebox{0.5\textwidth}{!}{\begin{tikzpicture}
\def\fsize{\large}
\pgfplotsset{every y tick label/.append style={font=\fsize}}
\pgfplotsset{every x tick label/.append style={font=\fsize}}

\begin{semilogyaxis}[%
width=4in,
height=3in,
scale only axis,
every outer x axis line/.append style={white!15!black},
every x tick label/.append style={font=\color{white!15!black}},
xmin=-16,
xmax=2,
xtick = {-16,-14,...,2},
xlabel={\fsize SNR (dB)},
xmajorgrids,
every outer y axis line/.append style={white!15!black},
every y tick label/.append style={font=\color{white!15!black}},
ymin=0.008,
ymax=1,
yminorgrids = true,
ylabel={\fsize Error probability $P_{\mathrm{e}}$},
ymajorgrids,
legend style={at={(1,0.93)},anchor=north east, draw=black,fill=white,legend cell align=left,font=\small}
]

\addplot [color=mycolor1,solid,line width=2.0pt,mark size=3pt,mark=o,mark options={solid}]
  table[row sep=crcr]{
-10	1\\
-8	1\\
-6	1\\
-4	0.04\\
-2	0.03\\
0     0.02 \\
2     0.015\\
};
\addlegendentry{S-OFDM, K=10, N=3935 \cite{chen2017sparse}};

\addplot [color=black,solid,line width=2.0pt,mark size=3pt,mark=*,mark options={solid}]
  table[row sep=crcr]{
-10	1\\
-8	1\\
-6	1\\
-4	0.43\\
-2	0.08\\
0	0.023\\
2    0.0016\\
};
\addlegendentry{Proposed Scheme, K=10, N=720};

\addplot [color=mycolor4,solid,line width=2.0pt,mark size=3pt,mark=triangle,mark options={solid,rotate=90}]
  table[row sep=crcr]{
-16 1\\
-14 0.9\\
-12 0.08\\
-10	0.078\\
-8	0.034\\
-6	0.02\\
-4	0.011\\
};
\addlegendentry{S-OFDM, K=100, N=29450 \cite{chen2017sparse}};

\addplot [color=mycolor3,solid,line width=2.0pt,mark size=3pt,mark=+,mark options={solid}]
  table[row sep=crcr]{
-16 1\\
-14 1\\
-12 1\\
-10	1\\
-8	0.89\\
-6	0.26\\
-4	0.008\\
};
\addlegendentry{Proposed Scheme, K=100, N=8600};

\end{semilogyaxis}
\end{tikzpicture}%
}
\caption{The error probability of asynchronous neighbor discovery is compared in this graph for fading model~I.}
\label{fig:sim_results2}
\end{figure}
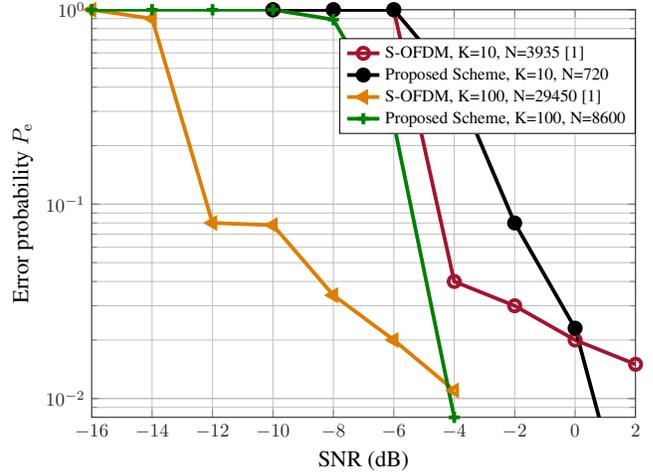 
Figure~\ref{fig:sim_results2} illustrates the performance of asynchronous neighbor discovery. The maximum transmission delay is set to $\mathcal{T}=20$ symbols, and the delay $\tau_k$ of active device~$k$ is drawn uniformly at random from $[0:\mathcal{T}]$. Again, we see a substantial performance improvement for the proposed scheme, compared to~\cite{chen2017sparse}. A given error probability can be achieved with a reduction in codeword length of approximately $82\%$ for $K = 10$, and $70\%$ for $K = 100$.

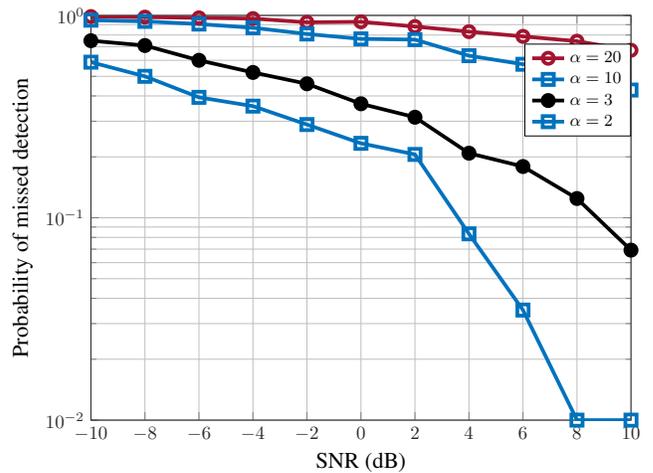
\begin{figure}[tbh]
\centering
\resizebox{0.5\textwidth}{!}{\begin{tikzpicture}
\def\fsize{\large}
\pgfplotsset{every y tick label/.append style={font=\fsize}}
\pgfplotsset{every x tick label/.append style={font=\fsize}}

\begin{semilogyaxis}[%
width=4in,
height=3in,
scale only axis,
every outer x axis line/.append style={white!15!black},
every x tick label/.append style={font=\color{white!15!black}},
xmin=-10,
xmax=10,
xtick = {-10,-8,...,10},
xlabel={\fsize SNR (dB)},
xmajorgrids,
every outer y axis line/.append style={white!15!black},
every y tick label/.append style={font=\color{white!15!black}},
ymin=0.01,
ymax=1,
yminorgrids = true,
ylabel={\fsize Probability of missed detection},
ymajorgrids,
legend style={at={(1,0.93)},anchor=north east, draw=black,fill=white,legend cell align=left,font=\small}
]

\addplot [color=mycolor1,solid,line width=2.0pt,mark size=3pt,mark=o,mark options={solid}]
  table[row sep=crcr]{
-10	0.9871\\
-8  0.9806\\
-6	0.9726\\
-4	0.9645\\
-2	0.9242\\
0 0.9300\\
2 0.8812\\
4 0.8313\\
6 0.7879\\
8 0.7457\\
10 0.6729\\
};
\addlegendentry{$\alpha=20$};

\addplot [color=mycolor2,solid,line width=2.0pt,mark size=3pt,mark=square,mark options={solid}]
  table[row sep=crcr]{
-10	0.9469\\
-8	0.9344\\
-6	0.9063\\
-4	0.8688\\
-2	0.8094\\
0 0.7656\\
2 0.7594\\
4 0.6324\\
6 0.5735\\
8 0.5861\\
10 0.4278\\
};
\addlegendentry{$\alpha=10$};

\addplot [color=black,solid,line width=2.0pt,mark size=3pt,mark=*,mark options={solid}]
  table[row sep=crcr]{
-10	0.75\\
-8	0.71\\
-6	0.6\\
-4	0.5226\\
-2	0.4593\\
0 0.3659\\
2 0.3140\\
4 0.2085\\
6 0.1792\\
8 0.1245\\
10 0.0692\\
};
\addlegendentry{$\alpha=3$};

\addplot [color=mycolor2,solid,line width=2.0pt,mark size=3pt,mark=square,mark options={solid}]
  table[row sep=crcr]{
-10	0.5875\\
-8	0.5\\
-6	0.3937\\
-4	0.3563\\
-2	0.2889\\
0 0.2333\\
2 0.2056\\
4 0.0833\\
6 0.035\\
8 0.01\\
10 0.01\\
};
\addlegendentry{$\alpha=2$};

\end{semilogyaxis}
\end{tikzpicture}%
}
\caption{This plot showcases the feasibility of leveraging fading coefficients to stitch sub-blocks together.}
\label{fig:sim_results3}
\end{figure}

\noindent
\textbf{\textit{Fading Model II:}} We consider a network with $K_{\mathrm{tot}}=2^{38}$, $K = 20$ and $\mathcal{T}=0$. The number of sub-blocks $n=4$, each sub-block is of length 12 and the parity-check bits are allocated as $[l_0~l_1~l_2~l_3]=[0~ 0~ 0~ 10]$. The scale parameter of Pareto distribution is set to $\eta=0.05$. Since the Pareto distribution has a heavy tail, the channel gains are distributed over a larger range and their estimates serve as an outer code in this model. Figure~\ref{fig:sim_results3} shows the probability of missed detection for various values of the shape parameter $\alpha$. It can be seen that increasing $\alpha$ worsens the performance, since the channel gains tend to concentrate within a small range.
Note that these are not comparison curves in this case because this fading scenario was not treated in~\cite{chen2017sparse}.


\bibliographystyle{IEEEbib}
\bibliography{IEEEabrv,MACcollision}

\end{document}